\documentclass{ws-procs975x65}

\begin{document}

\title{Stability of the Einstein static universe in modified theories of gravity}

\author{C. G. B\"ohmer}

\address{
Department of Mathematics and Institute of Origins, 
University College London,\\
Gower Street, London, WC1E 6BT, United Kingdom\\
\email{c.boehmer@ucl.ac.uk}
}

\author{L. Hollenstein}

\address{
D\'{e}partement de Physique Th\'{e}orique, Universit\'{e} de Gen\`{e}ve\\
24, Quai Ernest Ansermet, 1211 Gen\`{e}ve 4, Switzerland\\
\email{lukas.hollenstein@unige.ch}
}

\author{F. S. N. Lobo}

\address{
Centro de F\'{i}sica Te\'{o}rica e Computacional, Faculdade de Ci\^{e}ncias da Universidade\\ de Lisboa, Avenida Professor Gama Pinto 2, P-1649-003 Lisboa, Portugal\\
\email{flobo@cii.fc.ul.pt}
}

\author{S. S. Seahra}

\address{
Department of Mathematics and Statistics, University of New Brunswick,\\ 
Fredericton, New Brunswick, E3B 5A3, Canada\\
\email{sseahra@unb.ca}
}

%\begin{abstract}
%This article explains how to use World Scientific's ws-procs975x65
%document class written in \LaTeX2e. This article was typeset using
%ws-procs975x65.cls and may be used as a template for your contribution.
%\end{abstract}
%
%\keywords{Style file; \LaTeX}

\bodymatter

During the last decade, various modified theories of gravity have become very popular, $f(R)$ gravity theory probably being the most studied one~\cite{Capozziello:2003tk,Carroll:2003wy}. Although most models that are in agreement with observations are very close to general relativity, we have developed a much deeper understanding of the theory we wanted to modify. During the year 2009 it was Ho\v{r}ava-Lifshitz gravity~\cite{Horava:2008ih,Horava:2009uw} that excited the scientific community with a new paper appearing on the subject area every other day. Ho\v{r}ava proposed a power counting renormalizable theory for (3+1)-dimensional quantum gravity, which reduces to Einstein gravity with a non-vanishing cosmological constant in IR, but possesses improved UV behaviors.

In this work, we explore the stability of the Einstein static universe in such modified theories of gravity, for this study in general relativity see~\cite{Barrow:2003ni}. This can be motivated from various points of view. From a cosmological viewpoint it is the possibility that the universe might have started out in an asymptotically Einstein static state, in the inflationary universe context~\cite{Ellis:2002we}. On the other hand, the Einstein cosmos has always been of great interest in various gravitational theories. In general relativity for instance, generalizations with non-constant pressure have been analyzed in~\cite{IN:1976,Boehmer:2003uz}. In the context of brane world models the Einstein static universe was investigated in~\cite{Gergely:2001tn}, while its generalization within Einstein-Cartan theory can be found in~\cite{Boehmer:2003iv}. Finally, in the context of loop quantum cosmology, we refer the reader to~\cite{Mulryne:2005ef,Parisi:2007kv}. For the Einstein static universe in modified Gauss-Bonnet gravity, see~\cite{Bohmer:2009fc}. Finally, stability of the Einstein static universe in Ho\v{r}ava-Lifshitz gravity was analyzed in~\cite{Wu:2009ah,Boehmer:2009yz}.

By analyzing a simple background model and its perturbations one can study many properties of the modified theory in a rather explicit fashion. Let us give one example to elucidate this point. When $f(R)$ gravity became popular, it was believed that modifications of general relativity cannot stabilize solutions. However, as we showed explicitly in~\cite{Boehmer:2007tr}, this is not true and one can construct situations where the Einstein static universe, for instance, is stable with respect to a homogeneous perturbation. As we further showed~\cite{Seahra:2009ft}, this result in fact holds for all non-degenerate $f(R)$ ($f''(R) \neq 0$ etc) theories, see also~\cite{Goswami:2008fs}. Generic stability results in $f(R)$ gravity have been know since 1983~\cite{Barrow:1983rx}.

The mathematics required to perform these investigations is straightforward and well understood, namely linear perturbation theory. The principal idea is to express all quantities $u^i(x)$ in the form
\begin{equation}
  u^i(x) = u^i_{\rm bg}(x) + \varepsilon\, u^i_1(x) + \varepsilon^2\, u^i_2(x) + \ldots, 
  \label{eqn1}
\end{equation}
where the $u^i_{\rm bg}(x)$ describe a known exact solution of the field equations. It should be noted that there are situations where knowledge of $u^i_{\rm bg}(x)$ is not even necessary to solve to perturbed equations. In case of the Einstein static universe $x$ would be the cosmological time $t$, $u^i(x) = \{a(t),\rho(t),p(t)\}$, with $a(t)$, $\rho(t)$ and $p(t)$ being the scale factor, the energy density and the pressure respectively. The background solution would be a static solution of the field equations which effectively reduce to algebraic equations that determine the background values. In general, we can assume the field equations to take the form
\begin{equation}
  F^i(u^i,\nabla_\alpha u^i,\ldots, \nabla_\alpha \nabla_\beta \nabla_\gamma \nabla_\delta u^i) = 0.
  \label{eqn2}
\end{equation}
Here we restrict ourselves to theories which contain up to four derivatives. We also assume that the number of field equations matches the numbers of unknown functions, we exclude over-determined and under-determined systems. Notice that the field equations sourced by a perfect fluid are under-determined as long as no equation of state is assumed. Therefore, we consider a perfect fluid (no anisotropic stresses) with a linear barotropic equation of state, $p_{\rm bg}(\rho_{\rm bg})=w \rho_{\rm bg}$. Now, if $p(\rho)=w \rho$ is also assumed to hold at perturbative level then the sound speed of adiabatic pressure perturbations is given as $c_s^2=w$. Note that it would be interesting to investigate the situation when this assumption is dropped, $c_s^2 \neq w$. 

The approach described covers usual General Relativity and thus the entire field of cosmological perturbation theory, $f(R)$ gravity, modified Gauss-Bonnet gravity, or $f(\mathcal{G})$ gravity, and also Ho\v{r}ava-Lifshitz gravity. Note that the above treatment is sufficiently general to also include brane world models which pose additional technical challenges~\cite{Kodama:2000fa,Koyama:2000cc}.

In order to investigate the behavior of the perturbation, one now inserts Eq.~(\ref{eqn1}) into the field equations~(\ref{eqn2}) and linearizes the equations with respect to $\varepsilon$ (Taylor expansion about $\varepsilon=0$). In full generality, the field equations then become
\begin{multline}
  F^i(u^i_{\rm bg},\nabla_\alpha u^i_{\rm bg},\ldots, \nabla_\alpha \nabla_\beta \nabla_\gamma \nabla_\delta u^i_{\rm bg}) \\ + \varepsilon \left(\frac{\partial F^i}{\partial u^i}\right)_{\rm bg} u^i_1 + \ldots + \varepsilon \left(\frac{\partial F^i}{\partial \nabla_\alpha \nabla_\beta \nabla_\gamma \nabla_\delta u^i}\right)_{\rm bg}\nabla_\alpha \nabla_\beta \nabla_\gamma \nabla_\delta u_1^i  + O(\varepsilon^2) = 0.
  \label{eqn3}
\end{multline}
Those equations are now linear in the perturbed variables. In Gauss-Bonnet gravity, for example, this equation takes the simple form
\begin{equation}
  24 \kappa^3 \rho_{\rm bg}^2 (1+w)^2 f''(0)\, a_1''''(t) +
  2 a_1''(t) - \kappa \rho_{\rm bg} (1+w)(1+3w)\, a_1(t) = 0.
\end{equation}
In this equation $a_1$ is the perturbed scale factor and $f$ is the function which determines the modifications of the Gauss-Bonnet term. 

In general one finds a set of linear, coupled differential equations. Equations of this type can always be solved analytically. We can therefore conclude that the additional degrees of freedom in any modified gravity model lead to enhanced regions of stability in the parameter space.

\section*{Acknowledgments}
We would like to thank Peter Dunsby, Naureen Goheer, Roy Maartens and Luca Parisi for useful discussions. LH is supported by the Swiss National Science Foundation.

\end{document}